\documentclass[aip,reprint]{revtex4-1}
\usepackage{graphicx,amsmath,amssymb}

\begin{document}

\title{Ion-sensitive phase transitions driven by Debye-H\"uckel non-ideality}

\author{Kyle J.~Welch$^\dagger$}

\author{Fred Gittes$^*$}

\affiliation{$^1$Department of Physics \& Astronomy, Washington State University,
Pullman, WA 99164-2814. 
$^\dagger$Current address: Department of Physics, University of Oregon,
Eugene, OR 97403-1274.
$^*$Corresponding author.
}


\begin{abstract}
We find that the Debye-H\"uckel nonideality of dilute
aqueous electrolytes is sufficient to drive volume phase
transitions and criticality, even in the absence of a
self-attracting or elastic network.  Our result follows from
a Landau mean-field theory for a system of confined ions in
an external solution of mixed-valence counterions, where
the ratio of squared monovalent to divalent ion concentration
provides a temperature-like variable for the phase
transition.  Our analysis was motivated by long-studied
volume phase transitions via ion exchange in ionic gels, but
our findings agree with existing theory for
volume-temperature phase transitions in charged hard-sphere
models and other systems by Fisher and Levin, and McGahay
and Tomozawa.  Our mean-field model predicts a continuous
line of gas-liquid-type critical points connecting a purely
monovalent, divalent-sensitive critical point at one extreme
with a divalent, monovalent-sensitive critical point at the
other; an alternative representation of the Landau
functional handles this second limit.  It follows that
critical sensitivity to ion valence is tunable to any
desired valence ratio.  The critical or discontinuous
dependent variable can be the confinement volume;
alternatively the internal electrical potential may be more
convenient in applications.  Our simplified conditions for
ionic phase transitions to occur, together with our
relatively simple theory to describe them, may facilitate
exploration of tunable critical sensitivity in areas such as
ion detection technology, biological switches and osmotic
control.

\end{abstract}


\maketitle

\section{Introduction}
\label{Introduction}

Phase transitions in charged networks, as manifested by dramatic and reversible
swelling of a polymer gel, have been studied for over thirty years.
\cite{Tanaka-1980,Tanaka-1985,Fernandez-2001} In such gels expansion is driven
by the osmotic pressure of mobile counterions that are free to exchange with
an external ion population in a Donnan equilibrium.
\cite{Fernandez-2001,Ricka-1985} Osmotic pressure varies under exchange of
divalent for monovalent ions, so that that such systems can be critically
sensitive to---among other things---the relative external concentrations of
multivalent ions.  Accordingly, such phase transitions have long been discussed
with an eye to applications such as switches, artificial
muscles\cite{Tanaka-1982} and metal-ion detection,\cite{Jackson-1997} or as
candidate mechanisms for essential biophysical processes. \cite{Tasaki,Verdugo}
As in familiar liquid-vapor transitions, some element of self-attraction is
always required to drive the transition; in gels, effective self-attractions
are known to arise from relatively complicated network effects as theoretically
described by Flory.\cite{Flory-1953} In fact, such network behavior is by
itself rich enough to yield critical behavior with no ions present.
\cite{Fernandez-2001,Tanaka-1977}

Seeking a simplified framework for critical ionic sensitivity such as that
observed in charged networks, we have recast the theory of ionic phase
transitions into a solvable mean-field formulation, where we included in the
theory a nonideality of Debye-H\"uckel type, which is correct in the dilute
limit for all ionic solutions.  We found that in principle a self-attracting
network is not necessary for a discontinuous phase transition in the presence
of the power-law nonideality, which itself acts as an effective self-attraction
mediating an ionically driven phase transition of the gas-liquid type.  Our
model requires us to include a self-repulsive term to avoid runaway collapse
due to the non-ideal term.  In biological applications, intracellular charged
proteins may play the role of confined charges, and steric exclusion among
them would naturally provide the self-repulsive term.  In
fact\cite{Welch-Gittes-BPJ} charges on intracellular proteins in cells are
present in densities such that Debye-H\"uckel criticality could play a role
in osmotic control within organisms.

Within our mean-field theory, we find that by changing the magnitude of our
self-repulsive term, we can move the system through a line of critical points.
In this way, criticality is in principle tunable to occur at
any desired value of external divalent ion fraction.

Our finding of Debye-H\"uckel criticality was made in the context of ion
exchange in aqueous systems,\cite{Welch-Gittes-arXiv} but it is in accord
with established work in the context of volume-temperature transitions in
charged hard-sphere fluids initiated by Fisher and
Levin,\cite{Fisher-Levin-1993,Levin-1994} in semiconductor electron-hole
fluids, glasses, and molten salts by McGahay and Tomozawa\cite{McGahay-1992}
and in neutral polyampholytes (overall-neutral charged polymers) by Barbosa
and Levin.\cite{Barbosa-Levin}

Both our simplified formulation
of ionic transitions and our prediction of criticality in aqueous systems
much simpler than gel networks may illuminate mechanisms of tunable critical
sensitivity to ion valence or concentration that could underlie, for example,
biophysical cellular functions such as homeostasis.  \cite{Welch-Gittes-BPJ}
The theory may also lend itself to engineering applications involving ion
detection.  For these purposes the internal electric potential $\Phi$, which
we calculate, might be a more convenient dependent parameter than volume in
applications to ion detection or in biophysical roles for critical ionic
sensitivity.

\section{Donnan equilibrium with nonideality}
\label{Donnan equilibrium with nonideality}

Consider a population of $N_0$ ions each of charge $q_0$ (of either sign, in
our treatment) that may be bound to a mechanical structure, such as a polymer
network, or otherwise confined within a volume permeable to counterions.
These are the conditions for Donnan equilibrium \cite{Friedman,Starzak} in
which osmotic pressure of excess counterions goes hand-in-hand with an internal
voltage $\Phi$, relative to outside, with the same sign as $q_0$.  Even for
$\Phi\ne 0$, neutrality holds to a good approximation whenever the voltage
drop $\Phi$ occurs only at the boundary, or more generally if $N_0 \gg
C\Phi/q_0$, where $C$ is the system capacitance.

Osmotic pressure is exerted via the electrical potential drop at the gel
boundary.  External concentrations of counterions may be much more dilute
than the same species within the confinement volume.  Nonetheless, small
changes in the relative concentrations of external ions of differing valence
can lead to large changes in internal counterion valence ratios and thus in
the internal osmotic pressure, as sketched in Fig.~1.  When effective
self-attraction is present in the confined-ion system, to be supplied in our
case by the non-ideality of the internal solution, phase transitions in volume
and a critical point analogous to that in a gas-liquid system can occur.

\begin{figure}
\includegraphics[width=1.0\columnwidth]
{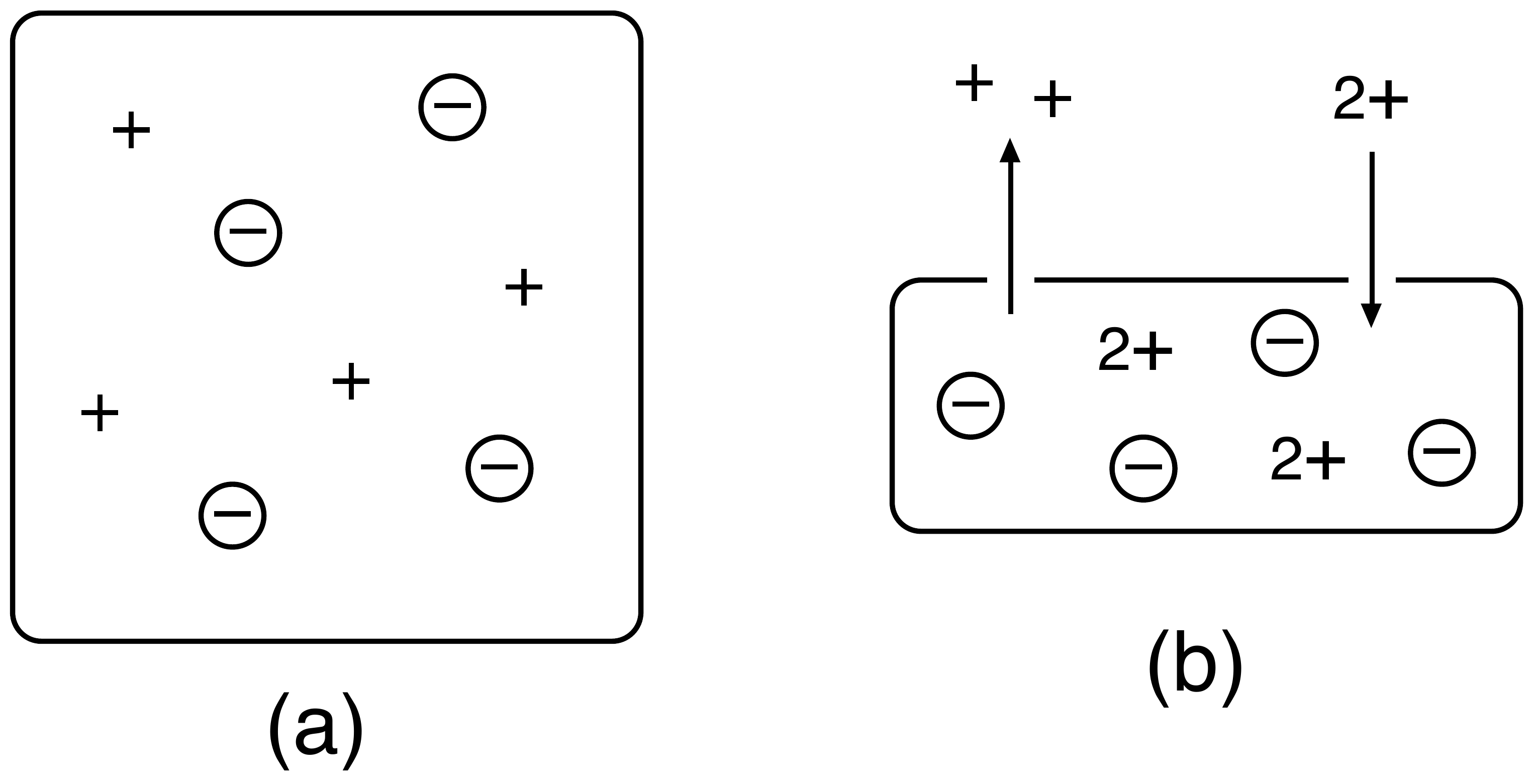}
\caption{
Schematic mechanism for a phase transition under changing relative
concentrations of mobile monovalent and divalent counterions; confined charges
are chosen negative in the diagram.  Divalent ions can exchange for monovalents,
reducing the internal osmotic pressure while preserving approximate neutrality.
An effective self-attraction of the confined-charge system can lead to a
volume phase transition, with external divalent fraction serving as an effective
temperature parameter.
}
\end{figure}

The immobile ions, confined within a variable volume $V$, have concentration
${c} = N_0/V = 1/v$.  Monovalent and divalent counterions with charges
$q_a=-q_0$ and $q_b=-2q_0$ are introduced at external concentrations $a$ and
$b$, and have within $V$ the concentrations ${a'}=N_a/V$ and ${b'}=N_b/V$.
Their free diffusion in and out is controlled by the the electrical potential
$\Phi$.  We introduce nonideality of mobile ions by way of a Debye-H\"uckel
interaction or correlation free energy \cite{Landau}
\begin{align}
    {F}_{\scriptscriptstyle{\text{DH}}}
    &\;=\;
    -
    {v_{\scriptscriptstyle\!B}^{\scriptscriptstyle 1/2}}\,
    {k_{\text{\tiny B}}T}\, 
    V 
    \Big(
    \sum_{\alpha}
    z^2_{\alpha} {c}_\alpha
    \Big)^{3/2}
\end{align}
where $q_\alpha = z_\alpha e$.  The essential features of
${F}_{\scriptscriptstyle{\text{DH}}}$ are that it is negative and contains
mobile-ion concentrations raised to the $3/2$ power.  The constant
$v_{\scriptscriptstyle B}$ is proportional to the cube of Bjerrum length.  In
standard aqueous conditions,
\begin{gather}
    \frac{1}{v_{\scriptscriptstyle B}}
    \;=\;
    (12\pi)^2
    \bigg(
    \frac{\epsilon\,{k_{\text{\tiny B}}T}}{e^2}
    \bigg)^{3}
    \;\approx\;
    3.4 {\,\mathrm{molar}}
    .
\end{gather}
To obtain closed-form solutions, we omit confined ions from the sum, and
replace the factor $( {a'} + 4{b'} )^{3/2}$ by $( {a'} + 2{b'} )^{3/2}$, which
will equal ${c}^{3/2}$ once neutrality is imposed.  Thus we employ the
Debye-H\"uckel-like interaction term
\begin{align}
    {F}_\text{int}
    &\;=\;
    -
    {v_{\scriptscriptstyle\!B}^{\scriptscriptstyle 1/2}}\,
    {k_{\text{\tiny B}}T}\, 
    V 
    ( {a'} + 2{b'} )^{3/2}
    .
\end{align}
With ${a'}=N_a/V$ and ${b'}=N_b/V$ and taking derivatives,
\begin{align}
    P_\text{int}
    & \;=\;
    -\frac{ \partial{F}_\text{int} }{\partial V}
    \;=\;
    -\tfrac{1}{2}
    {v_{\scriptscriptstyle\!B}^{\scriptscriptstyle 1/2}}
    \,{k_{\text{\tiny B}}T}\, 
    ( {a'} + 2{b'} )^{3/2}
    \\
    \frac{ \partial{F}_\text{int} }{\partial N_a}
    & \;=\;
    \,{k_{\text{\tiny B}}T} 
    \ln\gamma_a
    \;=\;
    -\tfrac{3}{2}
    \,{k_{\text{\tiny B}}T} 
    v_{\scriptscriptstyle B}^{\scriptscriptstyle 1/2}( {a'} + 2{b'} )^{1/2}
    \\
    \frac{ \partial{F}_\text{int} }{\partial N_b}
    & \;=\;
    \,{k_{\text{\tiny B}}T} 
    \ln\gamma_b
    \;=\;
    -3
    {v_{\scriptscriptstyle\!B}^{\scriptscriptstyle 1/2}}
    {k_{\text{\tiny B}}T}\, 
    ( {a'} + 2{b'} )^{1/2}
\end{align}
where activity coefficients $\gamma_a$ and $\gamma_b$ for the monovalent and
divalent ions are defined by \cite{Moore,Dickerson}
\begin{align}
    \mu_a 
    &\;=\;
    \mu_a^0 + {k_{\text{\tiny B}}T}\ln(\gamma_a a)
    .
\end{align}
Proportionality of $\ln\gamma$ to the square root of ionic concentration 
is the hallmark of Debye-H\"uckel behavior, a theory appropriate to
low ionic strength.\cite{Resibois,Horvath}
The free energy, with $\Phi$ externally controlled, is
\begin{multline}
    {F}(N_a,N_b,V,\Phi)
    =
    {F}_0(V)
    \, + \, 
    \\
    {F}_\text{int}(N_a,N_b,V) 
    \, + \, 
     (N_0\!-\! N_a\! -\! 2N_b )q_0\Phi
    \\
    \, + \, 
    N_a 
    {k_{\text{\tiny B}}T} 
    \Big[
    \ln\frac{N_a}{c_0 V} -1
    \Big]
    + 
    N_b 
    {k_{\text{\tiny B}}T} 
    \Big[
    \ln\frac{N_b}{c_0 V} -1
    \Big]
    .
\end{multline}
Here $F_0(V)$ describes mechanical constraints on the immobile ions, such as
a polymer network carrying the fixed ions or a membrane containing them.  The
logarithmic terms are the ideal free energy of the mobile ions, and $c_0$ is
a concentration scale which will cancel out.  The mechanical contribution to
pressure is $-\partial F_0 /\partial V = P_0$ (equivalently a function of $v$
or $c$).  The overall pressure and the chemical potentials $\mu_a=\partial
F/\partial N_a$ and $\mu_b=\partial F/\partial N_b$ are
\begin{align}
    P
    &\;=\;
    P_0(v)
    +({a'}+{b'}){k_{\text{\tiny B}}T}
    + P_\text{int}
    \\
    \mu_a
    &\;=\;
    -q_0\Phi + {k_{\text{\tiny B}}T}\ln\left(\frac{\gamma_a{a'}}{c_0}\right)
    \\
    \mu_b
    &\;=\;
    -2q_0\Phi + {k_{\text{\tiny B}}T}\ln\left(\frac{\gamma_b{b'}}{c_0}\right)
\end{align}
If we set $\mu_a = {k_{\text{\tiny B}}T}\ln(a/c_0)$ and $\mu_b = {k_{\text{\tiny
B}}T}\ln(b/c_0)$, $a$ and $b$ become effective concentrations, referred to
an ideal external solution.  To fix $\Phi$, we impose neutrality,
${c}={a'}+2{b'}$, and introduce a dimensionless potential 
\begin{gather}
    \phi 
    \;=\;
    q_0\Phi/{k_{\text{\tiny B}}T} 
\end{gather}
which will always be positive, since $\Phi$
will always be of the same sign as $q_0$.  We have altogether
\begin{gather}
    P
    \;=\;
    P_0(N_0/{c})
    +\tfrac{1}{2}({a'}+{c}){k_{\text{\tiny B}}T}
    -\tfrac{1}{2}
    {v_{\scriptscriptstyle\!B}^{\scriptscriptstyle 1/2}}\,
    {c}^{3/2}{k_{\text{\tiny B}}T}
    \\
    \phi \;=\;
    \ln(a'/a)
    -\tfrac{3}{2}
    {v_{\scriptscriptstyle\!B}^{\scriptscriptstyle 1/2}}\, 
    {c}^{1/2}
    \\
    2\phi \;=\;
    \ln(b'/b)
    -3 {v_{\scriptscriptstyle\!B}^{\scriptscriptstyle 1/2}}\, 
    {c}^{1/2}
\end{gather}
which, together with ${c}={a'}+2{b'} = N_0/v$, we will solve to find $P(v)$.
Combining the equations involving $\phi$, putting $2{b'}+{a'}-{c}= 0$ for
neutrality, and solving yields
\begin{align}
    {a'}
    &\;=\;
    \left(\frac{a^2}{4b}\right)
    \left[
    \left(1+ (8b/a^2){c}\right)^{1/2} - 1
    \right]
\end{align}
We define the dimensionless divalent parameter
\begin{align}
    \beta
    &\;=\;
    \frac{8b}{v_{\scriptscriptstyle B}a^2}
    .
\end{align}
To motivate the ratio $\beta \sim b/a^2$ by a chemical analogy, imagine $N_0/2$
divalent counterions $\mathrm{B}^{2+}$, complexed with an $N_0$-valent entity
$\mathrm{C}^{N_0+}$, cooperatively exchanging with $N_0$ monovalent counterions
$\mathrm{A}^{+}$ according to
\begin{align}
    \mathrm{CA}_{N_0} + (N_0/2)\mathrm{B}^{2+}
    \;\leftrightharpoons\;
    \mathrm{CB}_{N_0/2} + N_0\mathrm{A}^{+}
    .
\end{align}
Equilibrium, in the sense of mass action, then gives
\begin{align}
    \frac{[\mathrm{CB}_{N_0/2}]}{[\mathrm{CA}_{N_0}]}
    &\;=\;
    K\,\frac{[\mathrm{B}^{2+}]^{N_0/2}}{[\mathrm{A}^{+}]^{N_0}}
    \;=\;
    \text{const}
    \times
    \bigg(
    \frac{b}{a^2}
    \bigg)^{N_0/2}
    .
\end{align}
As $N_0\rightarrow\infty$, the monovalent-divalent exchange
$\mathrm{CB}_{n}\leftrightarrow \mathrm{CA}_{2n}$ becomes discrete at a
particular value of $b/a^2$.

\begin{figure}
\includegraphics[width=1.0\columnwidth]
{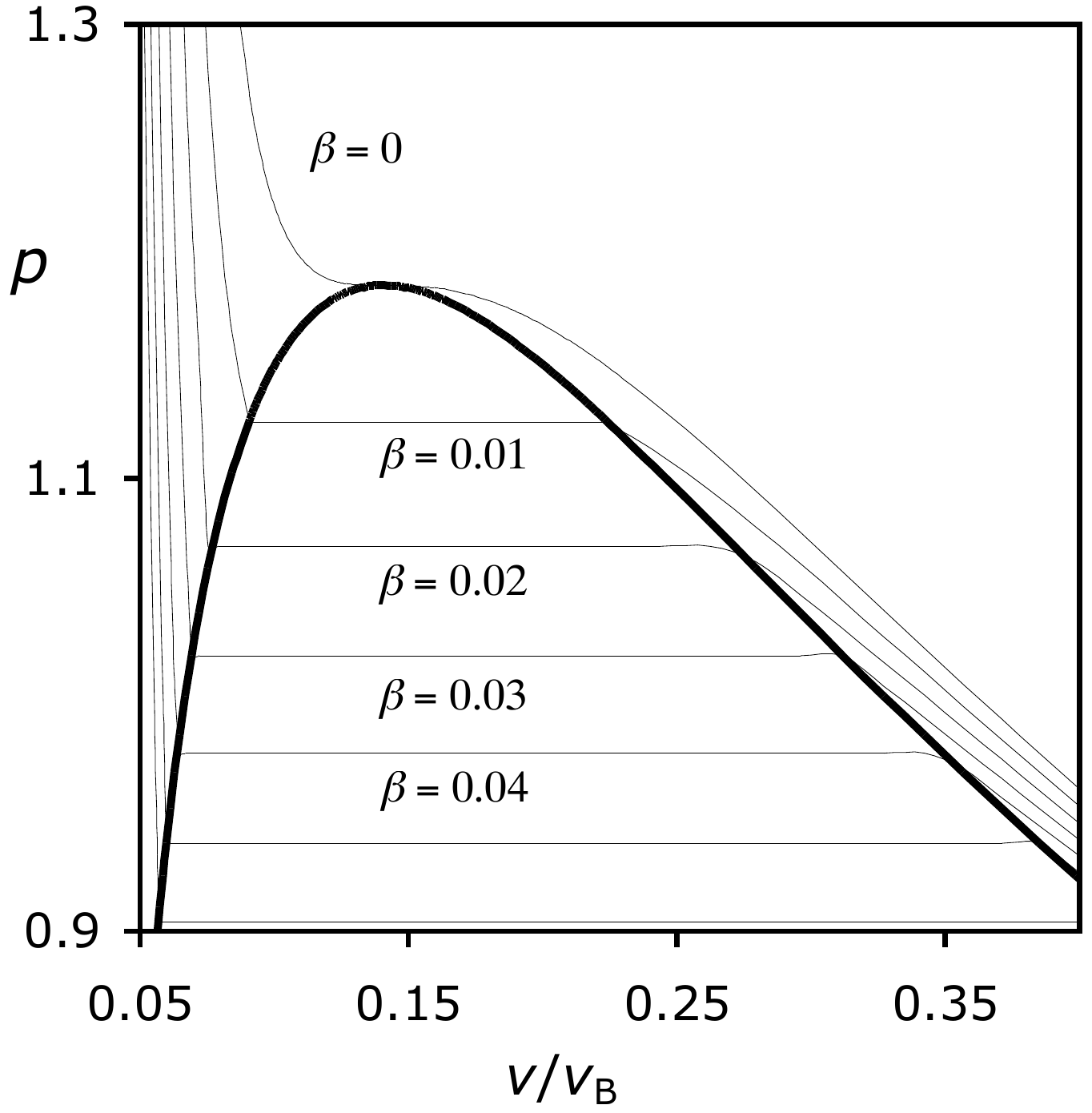}
\caption{
Coexistence diagram in dimensionless volume and pressure for a system tuned
by repulsion $\alpha$ to the monovalent ($\lambda=1$) critical point.  The
divalent parameter $\beta$ rises from zero upon addition of divalent ions,
so that $\beta<0$ is not possible for $\lambda=1$.  In similar coexistence diagrams for
$0<\lambda<1$, the upper region will be accessible.
}
\end{figure}

\section{Phase boundary}
\label{Phase boundary}

The dimensionless pressure
${p}(v) = P v_{\scriptscriptstyle B}/{k_{\text{\tiny B}}T}$ is
\begin{gather}
    {p}(v) 
    \;=\;
    {p_0}(v)
    +\frac{1}{2}\frac{v_{\scriptscriptstyle B}}{v}
    +
    \frac{1}{\beta}
    \bigg[
    \Big(1+ \beta\frac{v_{\scriptscriptstyle B}}{v}\Big)^{1/2} \!\! - 1
    \bigg]
    -\frac{1}{2}\Big(\frac{v_{\scriptscriptstyle B}}{v}\Big)^{3/2}
    \label{ptilde-of-v}
\end{gather}
where ${p_0} = P_0 v_{\scriptscriptstyle B}/{k_{\text{\tiny B}}T}$.  To make
the system stable against collapse to $v=0$ (due to the interaction term) we
require that ${p_0}(v)$ include a repulsion diverging faster than $1/v^{3/2}$.
We will use the minimal choice
\begin{align}
    {p_0}(v)
    &\;=\;
    \alpha \left(\frac{v_{\scriptscriptstyle B}}{v}\right)^2
    \;>\; 
    0
    \label{minimal-choice}
\end{align}
with $\alpha$ dimensionless.
The dimensionless electrical potential is
\begin{gather}
    \phi
    =
    \ln
    \frac{2}{\beta}
    \bigg[
    \Big(1 \!+\! \beta\frac{v_{\scriptscriptstyle B}}{v}\Big)^{1/2} \!\!\!\! - 1
    \bigg]
    -\frac{3}{2}
    \Big(\frac{v_{\scriptscriptstyle B}}{v}\Big)^{1/2}
    \!\!\!
    -\ln(v_{\scriptscriptstyle B}a )
    .
    \label{potential}
\end{gather}
From the Gibbs-Duhem relation (at constant $T$, or here $\beta$)
we obtain the chemical potential, linking coexisting $v$ and $v'$
at specified $\beta$, as 
\begin{align}
    \mu(v) 
    &\;=\;
    {p}v -\int {p}(v) dv
    \nonumber
    \\
    &\;=\;
    \frac{2\alpha}{v}
    -
    \frac{3}{2v^{1/2}}
    -\ln \Big[v \!+\! \sqrt{v(v\!+\!\beta)}\Big]
    +\tfrac{1}{2}.
\end{align}
Self-intersections of the curve $(p(v),\mu(v))$ yield phase boundaries
as in Figs.~2 and 3.

\section{Critical line}
\label{Critical line}

We solve our model in terms of the parameter
\begin{align}
    x &=
    \sqrt{\frac{v_{\scriptscriptstyle B}}{v}}
    .
\end{align}
At a critical point the conditions $p' (v) = p'' (v) = 0$ and $p' (x) = p''
(x) = 0$ are equivalent.  Regarding the critical value of $p$ as a function
of $x$ and $\beta$, and introducing the parameter $\lambda =
1/\sqrt{1\!+\!\beta_c x_c^2}$, we have the $x$-derivatives
\begin{align}
    {\tilde p}_{x}
    &\;=\;
    x_c\big[
    4\alpha \, x_c^2
    -\tfrac{3}{2}x_c
    +1
    \;
    +\lambda
    \big]
    \;=\;
    0
    \\
    {\tilde p}_{xx}
    &\;=\;
    12\alpha \, x_c^2
    -3x_c
    +1
    +\lambda^3
    \;=\;
    0
\end{align}
Eliminating $\alpha$ in favor of $\lambda$ gives
the line of critical points
\begin{align}
    p_c
    &\;=\; 
    x_c^2
    \big[
    \tfrac{1}{12} ( 1 - 6\lambda + \lambda^3 )
    + \lambda/(1+\lambda)
    \big]
    \label{pcrit}
    \\
    x_c
    &\;=\; 
    \tfrac{2}{3}
    [
    2
    +3\lambda
    -\lambda^3
    ]
    \label{xcrit}
    \\
    \alpha_c
    &\;=\;
    (1 +2\lambda -\lambda^3)/4x_c^2
    \label{alphacrit}
    \\
    \beta_c
    &\;=\;
    1/(x_c^2\lambda^2)
    \label{betacrit}
\end{align}
The interval $0 \le \lambda \le 1$ corresponds to $\infty\ge \beta \ge 0$.
Although we view $\alpha(\lambda)$ as the mechanism by which ionic critical
points can be tuned, we can also view $\alpha_c = \alpha(\lambda)$ as a
critical value of the repulsion strength.  With ionic conditions fixed, we
could drive transitions by modulating $\alpha$, analogous to transitions in
ionic gels induced by variation of solvent composition.\cite{Tanaka-1980}

\begin{figure}
\includegraphics[width=1.0\columnwidth]
{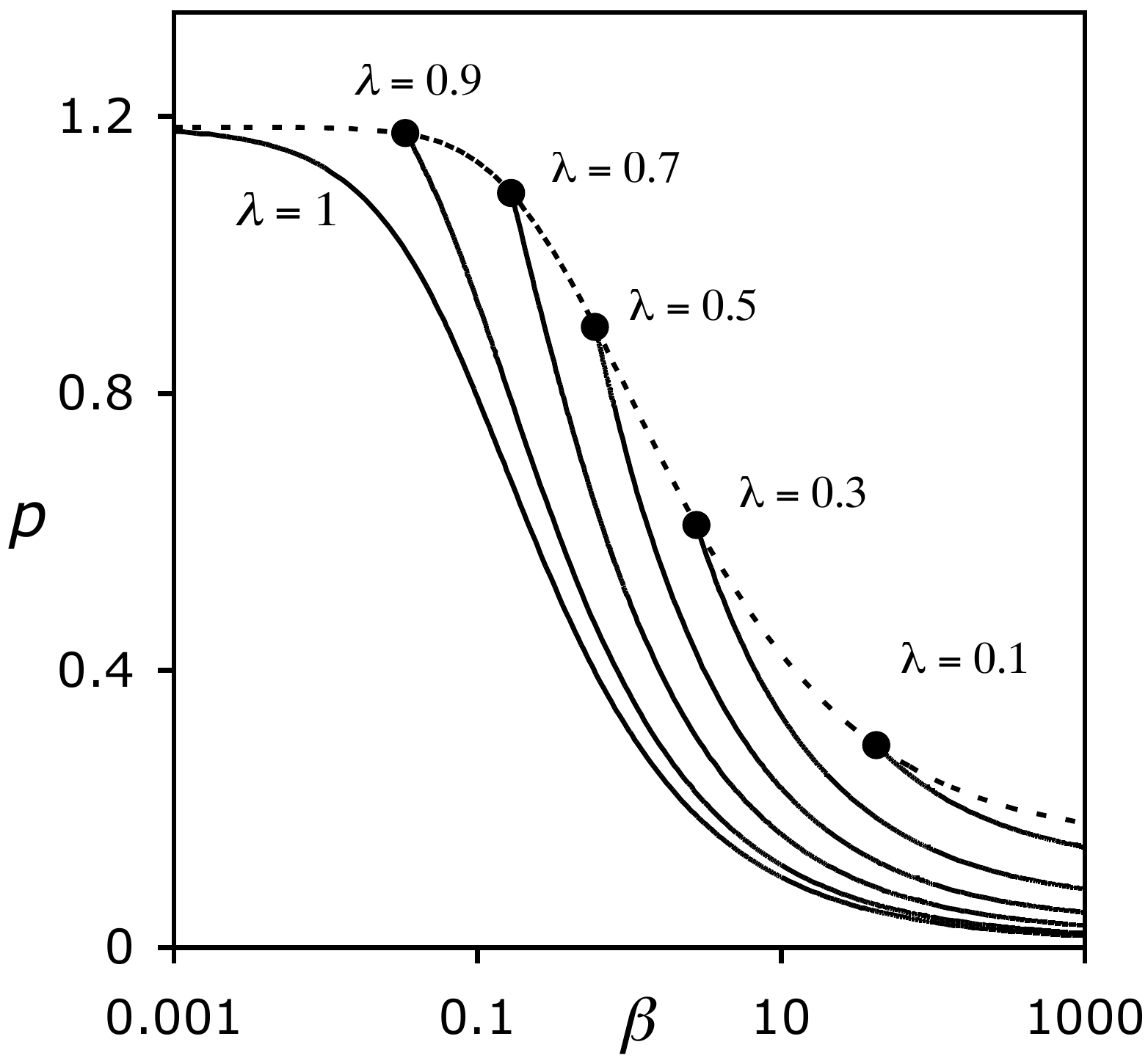}
\caption{
Critical line (dotted) connecting the monovalent ($\lambda=1$) and divalent
($\lambda=0$) critical points.  The divalent parameter  is $\beta$ and $p$
is dimensionless pressure.  Selected critical points and their phase boundaries
(solid) are shown.  The phase boundary becomes inaccessible (moving to $\beta
>\infty$) for the critical point at $\lambda=0$.
}
\end{figure}

About each critical point, with $\delta v = v-v_c$ and $\delta \beta
=\beta-\beta_c$, we construct a Landau expansion for the pressure,
\begin{align}
    p(\beta,\epsilon)
    &\;=\;
    p_c -  {A}\delta \beta + {B}\delta \beta\,\epsilon
    - {C}\,\epsilon^3
\end{align}
where $\epsilon =  \delta v + \kappa \delta \beta$ is an order parameter
linear in  $\delta v$ and $\delta \beta$.  In a potential application, we
imagine tuning a system to a critical point, $\beta=\beta_c$ and $p=p_c$.
Within mean-field theory, when the divalent ratio is changed by
$\delta\beta=\beta-\beta_c$ one predicts a singular expansion or contraction
\begin{align}
    \epsilon \approx
    -(A/C)^{1/3}
    \,
    (\delta\beta)^{1/3}
    \label{onethird}
\end{align}
where the cube root is taken with the same sign as $\delta\beta$.
As discussed below, this singular behavior in volume might in practice
be better monitored via the electric potential than the volume change
$\delta v$ or $\epsilon$.

To evaluate the Landau coeffients, we expand the pressure around the critical
point, to third order in $\delta v=v-v_c$ and to first order in $\delta\beta$,
giving
\begin{gather}
    p(v,\beta)
    \;\approx\;
    p_c
    +{\tilde p}_{\beta}\delta \beta
    +{\tilde p}_{v\beta}\delta \beta \epsilon
    +\tfrac{1}{6} {\tilde p}_{vvv}
    \epsilon^3
\end{gather}
where $\beta$ and $v$ subscripts denote partial derivatives and ${\tilde p}$,
${\tilde p}_{v}$, etc.\ are critical values. We have chosen $\kappa = {\tilde
p}_{vv\beta} /{\tilde p}_{vvv}$ to eliminate an $\epsilon^2$ term.  Evaluating
the various derivatives, the Landau parameters along the critical line are
\begin{align}
    {A}
    &\;=\;
    \tfrac{1}{2} x_c^4\lambda^3 / (1+\lambda)^2
    ,\quad
    {B}
    \;=\;
    \tfrac{1}{4}x_c^6 \lambda^3
    \\
    {C}
    &\;=\;
    \tfrac{1}{48}x_c^8
    \big[
    2+12\lambda-7\lambda^3+3\lambda^5
    \big]
    \\
    \kappa
    &\;=\;
    (\lambda^3/x_c)(\lambda^2+1)/(8 x_c\alpha_c - \beta_c\lambda^5)
\end{align}
where $x_c$ and $\alpha_c$ are known functions of $\lambda$ from
Eqs.~(\ref{xcrit}) and (\ref{alphacrit}).  $C$ has no zeros within $0 \le
\lambda \le 1$, while $\kappa$ is zero only at $\lambda = 0$.  At the monovalent
critical point ($\beta = 0$, i.e.\ $\lambda = 1$) Eqs.~(\ref{pcrit}) through
(\ref{betacrit}) yield $p_c = 32/27 \approx 1.185$, $x_c = 8/3 \approx 2.67$,
$\alpha_c  = 9/128 \approx 0.0703$, the Landau coefficients are
\begin{align}
    {A}
    &\;=\;
    2^{9}/3^4
    \;\approx\;
    6.321
    \\
    {B}
    &\;=\;
    2^{16}/3^6
    \;\approx\; 89.90
    \\
    {C}
    &\;=\;
    5(2^{21}/3^8)
    \;\approx\;
    1598.2
\end{align}
and the order parameter is $\epsilon = \delta v + (1/2)\delta\beta$.  Here
the system enters coexistence for any value of $\beta>0$, with $\epsilon
\approx -(0.158)\,\beta^{1/3}$.  A system tuned to the monovalent critical
point moves with infinite response towards a smaller volume with the
introduction of any divalents.  In the liquid-gas analogy, this path follows
the critical isobar.

At the purely divalent critical point ($\beta = \infty$, $\lambda = 0$) both
${A}$ and ${B}$ vanish, but we can here characterize the line of critical
points by the alternative parameterization
\begin{align}
    p(\beta,\epsilon)
    &\;=\;
    p_c 
    +{A'} \delta(\beta^{\scriptscriptstyle -1/2})
    -{B'} \delta(\beta^{\scriptscriptstyle -1/2})
    - {C}\,\epsilon^3
    \\
    {A'}
    &\;=\;
    x_c(1-\lambda)^{3/2}/(1+\lambda)^2
    \\
    {B'}
    &\;=\;
    \tfrac{1}{2}x_c^3
    (1-\lambda^2)^{3/2}
\end{align}
The parameter $\beta^{\scriptscriptstyle -1/2}$ is proportional to the
monovalent concentration at fixed divalent concentration.  The volume
singularity (\ref{onethird}) can be rewritten as
\begin{align}
    \epsilon \approx
    +\Big[
    \frac{{A'}}{{C}}
    \,\delta(\beta^{\scriptscriptstyle -1/2})
    \Big]^{1/3}
    \label{onethirdalt}
\end{align}
where again the cube root has the same sign as its argument.  At the divalent
critical point ($\beta = \infty$, $\lambda = 0$) Eqs.~(\ref{pcrit}) through
(\ref{betacrit}) yield $p_c = 4/27 \approx 0.1481$, $x_c = 4/3$, $\alpha_c =
9/64 \approx 0.1406$,
\begin{align}
    {A'}
    &=
    4/3
    ,
    \quad
    {B'}
    =
    32/27
    \;\approx\;
    1.185
    \\
    {C}
    &\;=\;
    2^{13}/3^9
    \;\approx\; 0.4162
\end{align}
For this divalent critical point at $\beta_c^{\scriptscriptstyle -1/2}=0$,
the order parameter is simply the volume, $\epsilon = \delta v$.  A purely
divalent system that is tuned to be critical moves with infinite response
towards a larger volume with the introduction of any monovalents, as $\delta
v \approx (1.214)\,\beta^{-1/6}$.

Returning to the dimensionless potential $\phi = q_0\Phi/{k_{\text{\tiny
B}}T}$ in Eq.~(\ref{potential}), along the critical line with $a$ constant
we find
\begin{align}
    \Big[
    \frac{\partial\phi}{\partial v}
    \Big]_c
    &\;=\;
    \frac{3}{4}
    \frac{x_c^3}{v_{\scriptscriptstyle B}}
    \Bigg[
    \frac{1+2\lambda - \lambda^3}
    {2+3\lambda - \lambda^3}
    \Bigg]
    \label{potentialderiv}
\end{align}
Eq.~(\ref{potentialderiv}) is nonzero along the entire line of critical points.
Since $\delta \phi\approx (\partial\phi/\partial v)_{c}\,\delta v$, the
potential will always exhibit the same power-law singularity as the volume.
In engineering applications, the internal electric potential may well be a
more convenient dependent parameter than volume. In a cell-biological context,
membrane electrical potential would be a likely route by which critical ion
sensing could be reported.

\section{Conclusions}
\label{Conclusions}

Our simplified conditions for criticality to occur in aqueous solutions, and
our mean-field framework for analyzing them, are intended to motivate
investigation of ionic phase transitions in ever simpler systems.  As a
fascinating analogue of liquid-gas transitions, and in the context of ionic
gels, such transitions have been a topic of discussion for many years.
Potential applications such as metal-ion detection and electrical sensitivity
\cite{Tanaka-1982,Jackson-1997} should remain applicable to phase transitions
as discussed here.  Of particular interest is our finding of a line of critical
points, such that ion-sensitive criticality is in principle tunable to occur
at any value of the divalent fraction.

We arrived at our finding of Debye-H\"uckel criticality\cite{Welch-Gittes-arXiv}
in the context of swelling transitions in ionic gels, a line of
experimental and theoretical work initiated by T.~Tanaka,
but subsequently found this conclusion
to be in accord with a distinct line of theory originating
with Fisher and Levin \cite{Fisher-Levin-1993} for temperature-density 
phase transitions in charged hard-sphere fluids, and of McGahay and Tomozawa
\cite{McGahay-1992} for transitions in electron-hole fluids in
semiconductors, glasses, and molten salts.  Subsequent theory by Barbosa and
Levin\cite{Barbosa-Levin} focused on neutral polyampholytes.  Our treatment of
nonideal Donnan equilibrium remains somewhat more phenomenological than that
of these workers, who employed ion-pairing refinements of
Bjerrum\cite{Bjerrum} and Fuoss\cite{Fuoss} to obtain quantitatively accurate
phase boundaries within their explicitly defined 
models.\cite{Fisher-Levin-1993} For simplicity, we have remained within a dilute
screening limit, even within the confined-ion region.

With regard to biological systems, we are applying the results
here to biological mechanisms\cite{Welch-Gittes-BPJ} such as
homeostasis (control of ion concentration) and cell volume
control, and in fact find that ionic levels in cells and charges
on intracellular proteins do fall within ranges appropriate for
Debye-H\"uckel criticality to play a role.

\vspace{2.0ex}

\begin{acknowledgments}

F.G.~thanks colleagues, and especially students in his biological
physics class at Washington State University for helpful and
enthusiastic conversations.  K.W.~especially thanks Professors D\"orte Blume
and J.~Thomas Dickinson for discussions of this material in the
context of an undergraduate thesis.  We are indebted to Yan Levin
for bringing to our attention existing lines of work on criticality in
Coulombic systems.

\end{acknowledgments}

\end{document}